\documentstyle[aps,preprint]{revtex}

\draft

\begin{document}
\title{Engineering Entanglement between two cavity modes }
\author{Manzoor Ikram$^{1}$and Farhan Saif$^{2}$}
\address{$^{1}$ Applied Physics Division, Pakistan Institute of Nuclear Science and\\
Technology, P. O. Nilore, Islamabad, Pakistan\\
$^2$Department of Electronics, Quaid-i-Azam University,
45320 Islamabad,
Pakistan.
}
\maketitle

\begin{abstract}
We present scheme for generation of entanglement between different modes of
radiation field inside high-Q superconducting cavities. Our scheme is based
on the interaction of a three-level atom with the cavity field for
pre-calculated interaction times with each mode. This work enables us to
generate complete set of Bell basis states and GHZ state.
\end{abstract}

\pacs{03.65.-w,03.67.-a,42.50.-p}

Quantum information theory implements quantum mechanics in to classical
information theory. The EPR-Bell correlations, and quantum entanglement \cite
{EPR,Bell}, in general, form the essential new ingredients, which
distinguish quantum information theory from its classical counterpart. An
entangled state of two or more quantum systems is a state which cannot be
factorized~\cite{Shamoni}. The most familiar example of an entangled state
is the Bohm state $\left( \left| \uparrow _{A},\downarrow _{B}\right\rangle
+\left| \downarrow _{A},\uparrow _{B}\right\rangle \right) /\sqrt{2},$
which, represents the state of two spin-1/2 particles decaying from a spin
zero parent \cite{Bohm}. The two particles are correlated, their spins are
always antiparallel, and they remain that way no matter what is the
separation between them. Hence entangled systems demonstrate non-local
quantum effects.

Entanglement is one of the main pillars of quantum compution \cite{Shor},
quantum cryptography \cite{Cryp}, quantum teleportation \cite{Telep} and
many other application of quantum information technology \cite
{Other,o2,o3,o4,o5,o6}, therefore, generation of entangled states and its
further applications are of immense importance. Several schemes have been
proposed for the generation of entangled states in atoms, ions, and photons
\cite{Atoms,a1,a2,a3,a4,a5,a6,Ions,Down}.

The first ever evidence of entangled state generation of two cavity fields
is seen in the teleportation procedure presented by Davidovich et.al. \cite
{David}. This entangled state of two cavity field of same mode occurs as an
intermediate step in the teleportation procedure and it requires presence of
one photon in either of the two cavities. The generation of GHZ state in two
cavities has been suggested in a similar manner \cite{Bergu}. An entangled
state in which number of photons between two cavities is fixed has also been
proposed \cite{Ikram}. All these schemes provide entanglement between
radiation fields of same mode in two cavities, recently, Ref.~\cite{Haroche} has
reported the existance of Bohm entangled state between different modes of
radiation field.

In this paper we propose schemes for the generation of entanglement between
different modes of electromagnetic field and engineer EPR-Bell basis and GHZ
entangled state~\cite{GHZ}. We propagate a three level atom through a cavity which
contains initially field modes in vacuum. We may express the three levels as
$|a\rangle $, $|b\rangle $ and $|c\rangle $ with their eigen energies as $%
E_{a}$ $E_{b}$ and $E_{c}$, as shown in Fig. 1. The dipole transition
between the upper two levels, $|a\rangle $ and $|b\rangle $, of the atom is
forbidden, whereas transitions from the two upper levels to lower level, $%
|c\rangle $ are allowed. We consider that frequencies, $\omega_A$ and $%
\omega_B$, of the two modes, $A$ and $B$, respectively, of cavity field are
in resonance with the transition frequencies, such that $\omega
_{A}=(E_{a}-E_{c})/\hbar $ and $\omega _{B}=(E_{b}-E_{c})/\hbar $. With the
help of a Ramsey field we prepare the upper two levels of the atom in linear
superposition before it enters the cavity field. We may express the initial
state of the system as,
\begin{equation}
\left| \phi (0)\right\rangle =\frac{1}{\sqrt{2}}\left[ |a\rangle +e^{i\phi
}|b\rangle \right] |0_{A},0_{B}>,
\end{equation}
where $\phi $ is the relative phase between two atomic states.

We write the interaction picture Hamiltonian in the dipole and rotating wave
approximation as
\begin{equation}
H=\hbar g_{1}\left( a\left| a\right\rangle \left\langle c\right| +a^{\dagger
}\left| c\right\rangle \left\langle a\right| \right) +\hbar g_{2}\left(
b\left| b\right\rangle \left\langle c\right| +b^{\dagger }\left|
c\right\rangle \left\langle b\right| \right)
\end{equation}
where $g_{1}$ and $g_{2}$ are vacuum Rabi frequencies of the two modes while
$a(a^{\dagger })$ and $b(b^{\dagger })$ are the annihilation(creation)
operators of the two cavity modes $A$ and $B$, respectively. The atom-field
state vector can be written as
\begin{equation}
\left| \psi _{at}^{(t)}(A,B)\right\rangle =C_{a,0,0}\left|
a,0,0\right\rangle +C_{b,0,0}\left| b,0,0\right\rangle +C_{c,1,0}\left|
c,1,0\right\rangle +C_{c,0,1}\left| c,0,1\right\rangle
\end{equation}
where $C_{a,n,m}$, $C_{b,n,m}$, and $C_{c,n,m}$ represent the time dependent
probability amplitudes for the atom to be in the states $\left|
a\right\rangle $, $\left| b\right\rangle $, and $\left| c\right\rangle $,
respectively, with $n$ number of photons in mode $A$ and $m$ photons in mode
$B$. The rate equations of these probability amplitudes can be obtained by
the Schr\"{o}dinger equation as
\begin{eqnarray}
\frac{d}{dt}C_{a,0,0} &=&-ig_{1}C_{c,1,0}, \\
\frac{d}{dt}C_{c,1,0} &=&-ig_{1}C_{a,0,0}, \\
\frac{d}{dt}C_{b,0,0} &=&-ig_{2}C_{c,0,1}, \\
\frac{d}{dt}C_{c,0,1} &=&-ig_{2}C_{b,0,0}.
\end{eqnarray}
Solving these differential equations in presence of the initial condition
mentioned in Eq.~(1), we find the atom field entangled state as
\begin{equation}
\left| \psi ^{(t)}(A,B)\right\rangle =\frac{1}{\sqrt{2}}\left[ \cos
g_{1}t\left| a,0,0\right\rangle -i\sin g_{1}t\left| c,1,0\right\rangle
+e^{i\phi }\cos g_{2}t\left| b,0,0\right\rangle -ie^{i\phi }\sin
g_{2}t\left| c,0,1\right\rangle \right]
\end{equation}
For the generation of maximally entangled field state between two cavities
such that if one cavity has one photon then the other will be in vacuum, the
atom after its interaction with the cavity fields, is required to be
detected in ground state $\left| c\right\rangle $. This leads to the
condition that probability amplitudes of the states $\left|
c,1,0\right\rangle $ and $\left| c,0,1\right\rangle $ are equal, i.e.,
\begin{equation}
\sin g_{1}t=\sin g_{2}t.
\end{equation}
The total probability of detecting the atom in ground state is determined as
\begin{equation}
P_{c}=\frac{1}{2}\left( \sin ^{2}g_{1}t+\sin ^{2}g_{2}t\right) .
\end{equation}
This probability becomes maximum when the time of interaction of atom with
mode $A$ and mode $B$ is $m\pi /2g_{1}$ and $n\pi /2g_{2}$, respectively.
Here, $m$ and $n$ are odd integer numbers. Hence, in order to generate two
mode entanglement the time of interaction of the atom with the cavity is odd
integer multiple of half of the Rabi cycle. This ensures that the cavity
will obtain one photon in either of the two modes when atom is detected in
ground state after its propagation through the cavity.

The interaction times of the atom with the two modes of the cavity field
would be different because of the different coupling constants of each mode
of radiation field. These interaction times of atom in the cavity can be
controlled by using a velocity selector before the cavity and then applying
Stark field adjustment so that atom becomes resonant with the cavity field
modes only for the suggested amount of time in each mode of the cavity field
\cite{David}.

Hence the atom passing in the superposition of levels $|a\rangle $ and $%
|b\rangle $ interacts with the two cavity modes $A$ and $B$ for $m\pi
/2g_{1} $ and $n\pi /2g_{2}$ interaction times, respectively. As a result
the atom leaves the cavity in ground state and develops an entangled state
between the two cavity modes, viz.,
\begin{equation}
\left| \psi (A,B)\right\rangle =\frac{-i}{\sqrt{2}}\left[ \left|
0_{A},1_{B}\right\rangle +e^{i\phi }\left| 1_{A},0_{B}\right\rangle \right] .
\end{equation}
In order to generate the other two Bell bases we control the interaction
time of the atoms with the cavities, hence we find
\begin{equation}
\left| \psi (A,B)\right\rangle =\frac{-i}{\sqrt{2}}\left[ \left|
0_{A},1_{B}\right\rangle -e^{i\phi }\left| 1_{A},0_{B}\right\rangle \right] .
\end{equation}

By using similar experimental setup we may prepare GHZ entangled states~\cite{GHZ},
between two different modes of the radiation field. We again consider a
three level atom in $V$ configuration, initially prepared in level $\left|
a\right\rangle $. We pass the atom through a system of two cavities which
are prepared initially in vacuum state. The transition from level $\left|
a\right\rangle $ to $\left| c\right\rangle $ is again in resonance with
cavity mode $A$, whereas the transition from $\left| b\right\rangle $ to $%
\left| c\right\rangle $ is in resonance with cavity mode $B$.

We adjust the interaction time of the atom with first cavity field such that
it sees a $\pi /2$ pulse. Hence, there occurs equal probability of finding
the atom in ground state $\left| c\right\rangle ,$ after contributing one
photon in the cavity mode $A,$ and of finding the atom in the excited state $%
\left| a\right\rangle ,$ leaving the cavity mode in vacuum state. As a
result we find an atom-field entanglement, such that,
\begin{equation}
\left| \psi_{at}(A,B)\right\rangle =\frac{1}{\sqrt{2}}\left[ \left|
a,0_{A}\right\rangle +\left| c,1_{A}\right\rangle \right] \otimes \left|
0_{B}\right\rangle .
\end{equation}

Before the atom enters in the next cavity, we apply a laser field resonant
to atomic transition $\left| b\right\rangle $ to $\left| c\right\rangle $.
The width of the beam is adjusted such that the exiting atom from the first
cavity field in the ground state $\left| c\right\rangle ,$ is pumped to
excited state $\left| b\right\rangle $ with unit probability. However, if
the exiting atom is in excited state $\left| a\right\rangle $ after
interacting with the cavity mode $A$, the laser field will provide no
excitation to the atom.

After passing through the laser field, the atom interacts with cavity mode $%
B $, which is initially in vacuum state. The interaction time of the atom
with the field is adjusted such that the atom in the excited state $\left|
b\right\rangle $ will be detected in ground state $\left| c\right\rangle $
with unit probability adding a photon in the cavity mode $B$. However, if
the atom enters the cavity in the excited state $\left| a\right\rangle $, it
will contribute no photon and will exit in the same atomic state, leaving
the cavity mode $B$ in the vacuum state. Therefore, we find the entanglement
of the two modes of radiation field with the atomic states as
\begin{equation}
\left| \psi_{at}(A, B)\right\rangle =\frac{1}{\sqrt{2}}\left[ \left| a,
0_{A},0_{B}\right\rangle +\left|c, 1_{A},1_{B}\right\rangle \right] .
\end{equation}
By taking projection over the atomic states, we may find the entangled state
between two cavity modes as
\begin{equation}
\left| \psi(A, B)\right\rangle =\frac{1}{\sqrt{2}}\left[ \left|
0_{A},0_{B}\right\rangle +\left|1_{A},1_{B}\right\rangle \right] .
\end{equation}

The interaction times of the atoms with the cavity field mode $A$, laser
field, and cavity field mode $B$ are found as $m\pi /4g_{1},$ $\pi /\Omega ,$
and $n\pi /2g_{2},$ respectively, where $m$ and $n$ are odd integers. {\rm %
Here }$\Omega ${\rm \ is the Rabi frequency of the laser field which
interacts with the atom between the two cavities.} If the relative
difference of interaction times of atoms with the two cavities is taken to
be $\pi $, we may generate the entangled state
\begin{equation}
\left| \psi (A,B)\right\rangle =\frac{1}{\sqrt{2}}\left[ \left|
0_{A},0_{B}\right\rangle -\left| 1_{A},1_{B}\right\rangle \right] .
\end{equation}
Hence, we can obtain the complete set of Bell basis by controlling the
interaction times of the atom with the cavities in both the schemes.

In order to generate a multi-mode entangled state we may repeat the same
process again as suggested above. We provide another laser pulse which is in
resonant to the transition from level $\left| c\right\rangle $ to another
higher level $\left| b_{1}\right\rangle .$ The atomic interaction with the
laser pulse occurs for a time $\pi /2\Omega _{1}$, where $\Omega _{1}$ is
the Rabi frequency of laser pulse. This pulse causes the atom to be in the
excited state $\left| b_{1}\right\rangle $ with unit probability. The atom,
in the excited state $\left| b_{1}\right\rangle $ then interacts with a
cavity mode $B_{1}$, initially in vacuum state, for an interaction time $%
m\pi /2g_{3},$ where $g_{3}$ is the vacuum Rabi frequency of $B_{1}$ mode of
the cavity field. Hence if the atom was in ground state $\left|
c\right\rangle $ after interacting with the second cavity then it will
contribute one photon to the cavity mode $B_{1}$ after this interaction,
whereas if it was in excited state $\left| a\right\rangle $ after second
cavity then the field will not interact with it because of the detuning. By
repeating the process using various different excited states we may develop
GHZ\ entangled state as
\begin{equation}
\left| \psi(A, B,....N) \right\rangle =\frac{1}{\sqrt{2}}\left[ \left|
0_{A},0_{B},....,0_{N}\right\rangle +\left|
1_{A},1_{B},....,1_{N}\right\rangle \right] ,
\end{equation}
which indicates entanglement between N-number of modes.

In order to measure any component of the Bell basis we may introduce a
standing wave field normal to the axis of the cavities providing entangled
state. A super cooled atom passing through the standing wave field and, thus
interacting with the entangled state would manifest interference pattern
unique to each of the Bell basis. A comparison of the interference pattern
with already stored patterns would reveal the information of entangled
states \cite{Farhan}.

In order to realize our suggested scheme in laboratory experiment within
microwave region, we may consider slow Rb atoms in higher Rydberg states which
have life time of the order of few milli seconds \cite{David}. These
slow atoms, initially pumped to high Rydberg state,
pass through a high-$Q$ superconducting cavity of dimension of a few
centimeters with a velocity of around 400 m/s~\cite{a1,David,Haroche}.
The interaction times of atom with cavities come out to be of the
order of few tens of microseconds which is far less than the cavity life
time. The high-$Q$ cavities of life time of the order of millisecond are
being used in recent experiments~\cite{Haroche}. The interaction time of the atom with
different cavities can be controlled by using a velocity selector and
applying Stark field adjustment in different cavities in order to make the
atom resonant with the field for right amount of time \cite{David}. The
atomic decay rates, interaction times, and cavity life time ensure that atom
does not decay spontaneously. As this entanglement remains only for the cavity
life time period so any application regarding this entangled state should be
accomplished during this period.

\begin{center}
{\bf Figure Captions}
\end{center}

Fig. 1. Schematic diagram for the generation of entangled state between two
cavity modes: We prepare three-level atom in superposition of upper two
levels by Ramsey field and let it pass through the cavity carrying two
different modes of radiation.

\end{document}